\documentclass[11pt]{article}
\usepackage{xspace,amssymb,amsmath,helvet,graphicx,url,authblk}
\usepackage[defaultcolor=red]{changes}
\usepackage{rotating}
\usepackage{amsmath}%
\usepackage{amsfonts}%
\usepackage{verbatim}
\usepackage[font={footnotesize}]{caption}
\usepackage{bm}
\usepackage{setspace}
\usepackage[cp1250]{inputenc}
\usepackage{fullpage}
\usepackage{booktabs}
\usepackage{multirow}
\usepackage{rotating}
\usepackage{pifont}
\usepackage{booktabs}
\usepackage{authblk}
\usepackage{setspace}
\usepackage{threeparttable}
\usepackage{tabularx}
\usepackage{pbox}
\usepackage{rotating}
\usepackage{subfig}
\begin{document}
\newcommand{\dee}{\,\mbox{d}}
\newcommand{\naive}{na\"{\i}ve }
\newcommand{\eg}{e.g.\xspace}
\newcommand{\ie}{i.e.\xspace}
\newcommand{\pdf}{pdf.\xspace}
\newcommand{\etc}{etc.\@\xspace}
\newcommand{\PhD}{Ph.D.\xspace}
\newcommand{\MSc}{M.Sc.\xspace}
\newcommand{\BA}{B.A.\xspace}
\newcommand{\MA}{M.A.\xspace}
\newcommand{\role}{r\^{o}le}
\newcommand{\signoff}{\hspace*{\fill} Rose Baker \today}
\newenvironment{entry}[1]%
{\begin{list}{}{\renewcommand{\makelabel}[1]{\textsf{##1:}\hfil}%
\settowidth{\labelwidth}{\textsf{#1:}}%
\setlength{\leftmargin}{\labelwidth}
\addtolength{\leftmargin}{\labelsep}
\setlength{\itemindent}{0pt}
}}%
{\end{list}}
\title{Psychometric Tests: Quantifying the Consequences of Low Reliability and Improving Reliability Estimation}
\author[1]{Rose Baker}
\affil[1]{School of Business, University of Salford, The Crescent, Salford M5 4WT, UK, email rose.baker@cantab.net}
\maketitle
\begin{abstract}
We introduce a set of consequence-based measures that quantify the 
misclassification arising from imperfect reliability in multi-item psychometric 
scales. Particular attention is given to the probability of individuals 
in extreme latent-trait percentiles being correctly identified from their 
observed scores. These cost functions provide a principled and interpretable 
way to characterise how inadequate reliability distorts classification 
and reduces the informational value of test scores.

We also examine methodological issues in estimating reliability under 
common-factor models, with emphasis on McDonald's omega and the construction 
of accurate confidence intervals. To address the computational burden 
of model fitting, especially in small samples, we derive a modified 
version of Cronbach's alpha that closely approximates omega when the 
common-factor model holds. This estimator has comparable sampling variability 
to omega while requiring no parameter estimation, offering a practical 
and computationally efficient alternative.

\end{abstract}
\section*{Keywords}
Multi-item scale, congeneric model, reliability, misclassification cost, Cronbach's alpha, 
McDonald's omega
\section{Introduction}
Psychometric multi-item scales are widespread, and are used not only 
directly in psychology, but also \eg in marketing to measure consumer 
satisfaction (see \eg Raykov and Marcoulides, 2011). Examination questions
that measure a student's ability in a subject area are another example.

The reliability of the scale is important; this is the proportion of 
the variance of test results that is due to variation in ability (latent 
score) rather than to statistical error. Traditionally, it was measured 
by Cronbach's alpha (\eg Cronbach, 1951). Feldt {\em et al} (1987) derived 
confidence intervals for alpha. Novick (1967) showed that Cronbach's alpha 
underestimates reliability except when the tau-equivalent model holds.

More recently, attention has largely switched to McDonald's omega as 
a measure of test reliability (McDonald, 1999). Zinbarg {\em et al} (2005) 
give a comparison of the two measures. McDonald's omega is thought to 
be better, except for small samples where estimation of the large number 
of parameters required becomes a problem for  omega, and the maximum-likelihood 
method can give unreliable results.

Flora (2020) explains how to use R to calculate McDonald's omega. There 
is however a lack of detailed information about the computation of an 
estimate of McDonald's omega and its associated confidence interval, which 
is a lack we address here. It is also possible to use the responses to 
the individual items more efficiently by weighting each response by its 
inverse variance. This is shown here to lead to a score with a higher 
reliability, where it can be applied. It cannot be used for examination
scores, as it would be unfair to weight up some questions more than others,
when students have been told that all questions carry the same maximum
mark.

Besides computing test reliability as accurately as possible,  one should
know what the consequences of low reliability are. Various measures of
the `cost' of low reliability are derived here. One that is computationally
least tractable but is most useful is the probability that an individual
whose true score would place them in the top (say) 5\% of scores, is 
indeed placed there. This statistic is unity at a reliability of unity, 
and decreases to 0.05 as the reliability goes to zero. It is often said 
that Cronbach's alpha should be at least 0.7, but the consequences of 
low reliability can be made more explicit than this.

Rather than fitting the common factor model, and bifactor models which 
require even more parameters, Cronbach's alpha reflects an earlier approach 
that does not require parameter estimation at all. This is an attractive 
proposition, with the drawback that the reliability is underestimated 
when the tau-equivalent model does not hold. Here, we add a factor loading 
variance term to Cronbach's alpha to rectify this deficiency, giving a 
quick and easy way to compute reliabilities when the common factor model 
holds.

Next, the derivation of some useful cost functions is given, followed 
by some useful measures of reliability and the computation of McDonald's omega. The maximum-likelihood estimation 
of McDonald's omega is then studied using simulation,and finally some fits 
to data exemplify the methodology.

\section{The effect of low reliability  on multi-item scale performance}
\subsection{What is reliability: the common-factor model and McDonald's omega}
With $k$ items, let the $j$th score for the $i$th of $n$ individuals 
be $x_{ij}$. Then the single-factor model assumes that $x_{ij}=\mu_j+\sigma_{j}\epsilon_{ij}+\lambda_j	\theta_i$,
where all random variables $\epsilon_{ij}$ and $\theta_i$ are independent
standard normal random variables, and $\theta_i \sim N[0,1]$ is the latent score for an individual. 
Hereafter we set $\mu_j=0$ with the
understanding that now $x_{ij} \rightarrow x_{ij}-\sum_{i=1}^n x_{ij}/n$.
The total score
\[T_i=\sum_{j=1}^k x_{ij}=\sum_{j=1}^k \sigma_j \epsilon_{ij}+(\sum_{j=1}^k \lambda_j)\theta_i.\]
and $\hat{\theta}_i=T_i/\sum_{j=1}^k \lambda_j$.
Hence $\hat{\theta}_i \sim N[\theta_i,A]$, where $A=\sum_{j=1}^k \sigma_j^2/(\sum_{j=1}^k
\lambda_j)^2$.

An advantage of this model is that the model form is invariant to aggregation
of items. That is, if we take the total of items 1 and 2, this is
\[T=x_{i1}+x_{i2}=\sigma_1\epsilon_{i1}+\sigma_2\epsilon_{i2}+(\lambda_1+\lambda_2)\theta_j.\]
The terms in $\epsilon$ sum to a normally-distributed random variable,
so the model form is unchanged. In practice, this property is needed,
as items are not identical, and could be regarded as made up of  variable
numbers of sub-items.

The reliability $R$ is defined as the proportion of the total variance
of $T_i$ that is due to the true score $\theta_i$. Hence
\begin{equation}R=\frac{(\sum_{j=1}^k  \lambda_j)^2}{\sum_{j=1}^k \sigma_j^2+(\sum_{j=1}^k  \lambda_j)^2}=1/(1+A).\label{eq:r}\end{equation}
This is McDonald's omega.

\subsection{Cost measures}
\subsubsection{The proportion correctly assigned to  top or bottom percentiles}
Identifying outstanding individuals, either with the highest or lowest
scores, is important, and so the probability that an individual whose
score $\theta_i$ places them in (say) the top 5\% of scores is a measure
of the performance of the multi-item scale. Let the top $p$ percentile
of $\theta$ be $p$, so that $1-\Phi(\theta_0)=\Phi(-\theta_0)=p$, where
$\Phi$ is the normal distribution function and $p$ is typically 0.01,
0.025 or 0.05. Allowing the true score $\theta$ to vary, $\hat{\theta}
\sim N[0,A+1]$, where as before $a=(1-r)/r$, so the variance of $\hat{\theta}$ is $1/R$. Hence the
p-percentile point of $\hat{\theta}$ occurs at $\beta$, where $\Phi(-\beta/\sqrt{R})=\Phi(-\theta_0)$
or $\beta=\theta_0/\sqrt{R}$.

Writing $f$ for pdfs (probability density functions), the probability we seek is
\[P=\frac{\int_\beta^\infty \dee \hat{\theta} \int_{\theta_0}^\infty f(\hat{\theta}|\theta)f(\theta)\dee \theta}{p},\]
\ie the probability that $\hat{\theta}$ is in the top percentile, given
that the true score is there.

The integral over $\theta$ can be done by completing the square in the
exponent, to yield the result
\begin{equation}P=\frac{\sqrt{R}}{\sqrt{2\pi}p}\int_{\theta_0/\sqrt{R}}^\infty \Phi(\frac{R\hat{\theta}-\theta_0}{\sqrt{1-R}})\exp(-\hat{\theta}^2R/2)\dee \hat{\theta}.\label{eq:cost}\end{equation}
This can be rewritten in terms of the bivariate normal distribution function,
but cannot be simplified in terms of elementary functions. It is an expensive
computation, because the integrand contains the special function $\Phi$.

\begin{figure}
\centering
\makebox{\includegraphics{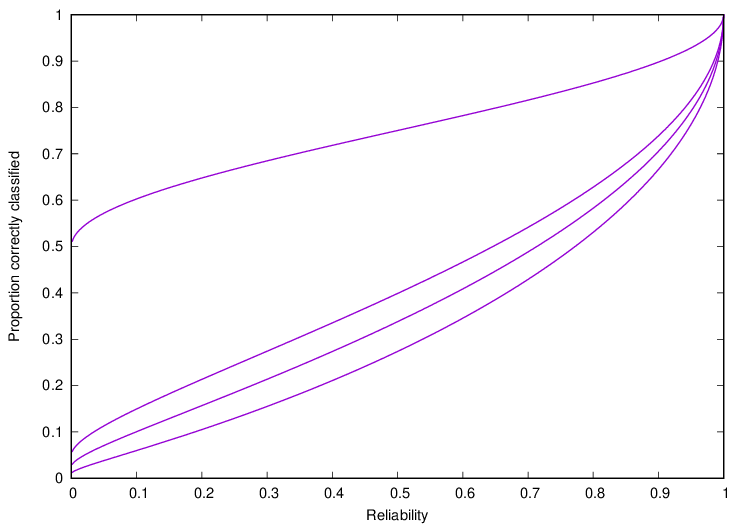}}
\caption{\label{figb}Proportion of individuals correctly classified into the top 50\%, 5\%, 2.5\% and 1\% by reliability.}
\end{figure}

Figure \ref{figb} shows a plot of $P$ against reliability $R$. It can
be seen that $P$ changes from unity ($\hat{\theta}=\theta$) when $R=1$
to $p$ when $R \rightarrow 0$. In that case, the test result gives no
information about $\theta$, and the probability that the test score $T$
is in the upper $p$ percentile is $p$.

In fact, $P$ is the power of a statistical test that $\theta_i$ is in
the top $p$ percentile. We normally require the power of statistical 
tests to be at least 0.8, so from figure \ref{figb} the reliability should 
be at least 0.9 for the test to be useful.

\subsubsection{The probability that the better individual is classed as better}
Given two individuals, suppose that $\theta_1 > \theta_2$. What is the
probability that the measured scores reflect this, \ie $\hat{\theta}_1
> \hat{\theta}_2$? We require $\hat{\theta}_1-\hat{\theta}_2 > 0$, and
$\hat{\theta}_1-\hat{\theta}_2 \sim N[\theta_1-\theta_2,2A]$, where $A=R/(1-R)$.
hence
\[\text{Prob}(\hat{\theta}_1-\hat{\theta}_2 > 0|\theta_1 > \theta_2)=\Phi\{\frac{(\theta_1-\theta_2)\sqrt{R})}{\sqrt{2(1-R)}}\}.\]
Integrating over the distribution of $\theta_1-\theta_2 > 0$, we have
that the required probability $P$ is $\text{Prob}=2\int_0^\infty \Phi(\sqrt{R}x)\phi(x)\dee
x$. It can be shown that $\int_0^\infty \Phi(ax)\phi(x)\dee x=1/4+\tan^{-1}(a)/2\pi$. 
This can be done by differentiating under the integral sign with respect 
to $a$, doing the resulting normal integral, and then integrating over 
$a$, or otherwise. Hence finally
\[P=1/2+\sin^{-1}(\sqrt{R})/\pi.\]
This is also the formula to which (\ref{eq:cost}) reduces when $\theta_0=0$,
\ie the probability that an individual is correctly classified as above
or below average, and so is shown in figure \ref{figb}. When $R=0$, $P=1/2$,
and when $R=1$, the arcsine is $\pi/2$ and  $P=1$.

\subsubsection{The overall cost of misclassification}
With true score $\theta$, the proportion of individuals having lower 
scores would be $\Phi(\theta)$. With an imperfect test, this becomes $\Phi(\theta+\sqrt{(1-R)/R}\epsilon)$.
The expectation
\[\int_{-\infty}^\infty \Phi(\theta+\sqrt{(1-R)/R}\epsilon) \phi(\epsilon)\dee \epsilon=\Phi(\theta\sqrt(R),\]
as can be readily proved. We see that the `placement' of an individual
tends towards the midpoint of $1/2$ as $R$ decreases. Since $\Phi(\theta\sqrt{R})
< \Phi(\theta)$, the average deviation from the true placement is
\[C=2\int_0^\infty (\Phi(\theta)-\Phi(\theta\sqrt{R}))\phi(\theta)\dee \theta.\]
This evaluates to
\[C=1/4-\tan^{-1}(\sqrt{R})/\pi.\]

\begin{figure}
\centering
\makebox{\includegraphics{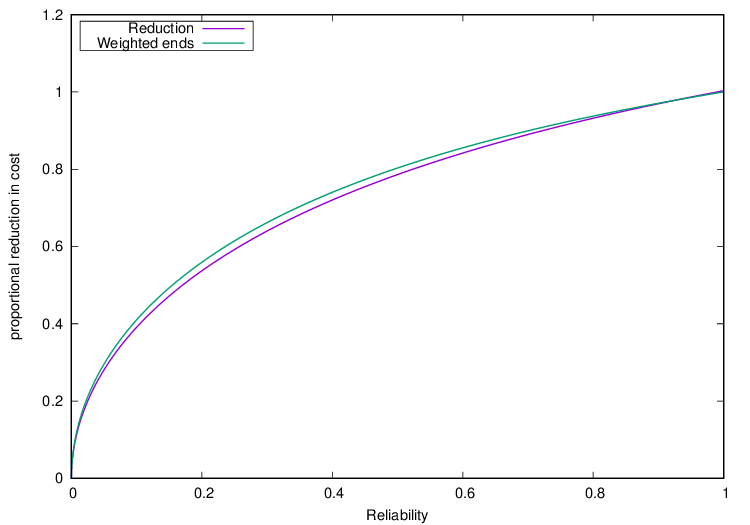}}
\caption{\label{figa}Reduction in proportional cost of misclassification versus reliability.}
\end{figure}

It is also possible to calculate the cost when weighting results at each
end of the scale by $\Phi(\theta)-1/2$, so that the cost decreases from
each end and is zero at the middle. Thus
\[C=2\int_0^\infty (\Phi(\theta)-\Phi(\sqrt{R}\theta))\phi(\theta)(\Phi(\theta)-1/2)\dee \theta.\]
In this case,  the cost evaluates to
\[C=1/6-\tan^{-1}(\sqrt{R/(2+R)})/\pi.\]
Figure \ref{figa} shows the proportional reduction in cost on moving 
from $R=0$ to $R=1$ for these two cases. The curves are quite similar.

Hopefully these cost measures will enable practitioners to understand
the consequences of low reliability.
\subsection{The probability that two tests agree}
So far, all cost measures have given insight into the effect of low reliability, 
but have not produced testable predictions.

Here we compare scores $\hat{\theta}_1$ and $\hat{\phi}_1$ for individual 
1 from two tests with reliabilities $R_1$ and $R_2$ with the corresponding 
scores for individual 2, and ask what is the probability that two random 
individuals have the same ranking under both tests.

Let the true abilities in the two tests be $\theta$ and $\phi$. Then 
$\hat{\theta}_1|\theta \sim N[\theta,(1-R_1)/R_1]$ and $\hat{\phi}_1|\phi 
\sim N[\phi,(1-R_2)/R_2]$, etc.

We have
\[\hat{\theta}_1-\hat{\theta}_2 \sim N[0,2(1-R_1)/R_1],\]
\[\hat{\phi}_1-\hat{\phi}_2 \sim N[0,2(1-R_2)/R_2].\]
We seek the probability
\[P=2\text{Prob}(\hat{\theta}_1 > \hat{\theta}_2, \hat{\phi}_1 > \hat{\phi}_2),\]
which is the probability that the two test results agree. We change variables 
to $u=(\theta-\phi)/\sqrt{2}$, $v=(\theta+\phi)/\sqrt{2}$. Then $P=\int_{-\infty}^\infty 
\Phi(u/\sqrt{A_1})\Phi(u/\sqrt{A_2})\phi(u)\dee u$. We have that
\[\int_{-\infty}^\infty \Phi(ax)\Phi(bx)\phi(x)\dee x=1/4+\sin^{-1}(\frac{ab}{\sqrt{1+a^2}\sqrt{1+b^2}}),\]
and so finally $P=1/2+\sin^{-1}(\sqrt{R_1 R_2})/\pi$.

\section{Different measures of reliability}
\subsection{Weighted scores and reliability}
Although the sum of item scores $T_i=\sum_{j=1}^k x_{ij}$ is commonly
used, a weighted score $W_i=\sum_{j=1}^k w_i x_{ij}$ has the least variable
mean if $w_i=1/\sigma_i^2$. Then the unbiased
estimator of $\theta$ is
\[\hat{\theta}=\frac{\sum_{j=1}^k\lambda_j x_j/\sigma_j^2}{\sum_{j=1}^k \lambda_j^2/\sigma_j^2}.\]
This has variance
\[\text{var}(\hat{\theta})=\frac{\sum_{j=1}^k \lambda_j^2/\sigma_j^2+1}{\sum_{j=1}^k \lambda_j^2/\sigma_j^2}.\]
The variance would be unity (that of the prior distribution of $\theta$)
if any of the variances were zero, so the reliability is
\[R_i=\frac{\sum_{j=1}^k \lambda_j^2/\sigma_j^2}{1+\sum_{j=1}^k \lambda_j^2/\sigma_j^2}.\]
One or more items having a large variance lowers $R$ drastically but 
merely means that the item has a small weight in $R_i$.

It seems from data fitting that $R_i > R$. For this to be always true,
we require that
\[\sum_{j=1}^k (\lambda_j/\sigma_j)^2\sum_{j=1}^k \sigma^2_j > \{\sum_{j=1}^k (\lambda_j/\sigma_j)^2\sigma_j^2\}^2.\]
but this is just the Cauchy-Schwartz inequality applied to the vectors
$\lambda_j/\sigma_j$ and $\sigma_j^2$. Hence indeed using the test results
like this has the advantage that $R_i > R$. For $R_i=R$ we require that
$\lambda_j \propto \sigma_j^2$. This occurs \eg when all $\lambda_j$ 
are equal, and all $\sigma_j$ are equal, the tau-equivalent model. Hence 
in general $R_i \ge R$.
\subsection{From omega to a modified alpha}
It is interesting to try to estimate reliability from (\ref{eq:r}) without 
fitting data to estimate parameters, in the spirit of Cronbach's alpha.
Let the sample covariance matrix of scores be ${\bf V}$.

Then  under the common-factor model
\[\sum_{i /= j}V_{ij}=(\sum_{j=1}^k \lambda_j)^2-\sum_{j=1}^k \lambda_j^2.\]
Squaring the elements of ${\bf V}$ to obtain the `Hadamard square', we 
have that $\sum_{i /= j}V_{ij}^2=\sum_{i /= j}\lambda_i^2\lambda_j^2$.
Imagine that the loadings $\lambda_i$ arise from a distribution of loadings 
with mean $\mu$ and variance $\tau^2$ and apply the method of moments.
Then $\text{E}\sum_{i /= j}V_{ij}^2=k(k-1)(\mu^2+\tau^2)^2$, so $\hat{\tau^2}=(\bar{c^2})^{1/2}-\bar{c}$, 
where we denote the average covariance and variance by $\bar{c}$ and $\bar{v}$ 
respectively.

Now $(\sum_{j=1}^k \lambda_j)^2 \simeq k^2 \bar{c}+k\hat{\tau^2}$ and 
the observed variances $s_j^2=\sigma_j^2+\mu^2+\hat{\tau}^2$, so finally (\ref{eq:r}) 
becomes
\begin{equation}\alpha_m=\frac{k\bar{c}+\hat{\tau^2}}{(k-1)\bar{c}+\bar{v}}\label {eq:alpham}\end{equation}

This is exactly Cronbach's alpha, but with the numerator increased by
the estimated variance of the loadings; note that McDonald (1999) finds 
that the difference between omega and alpha for the common-factor model 
is proportional to the variance of the factor loadings. Hence away from 
tau-equivalent models where $\tau^2=0$, this adjusted Cronbach's alpha
gives a  higher reliability.

\begin{figure}
\centering
\makebox{\includegraphics{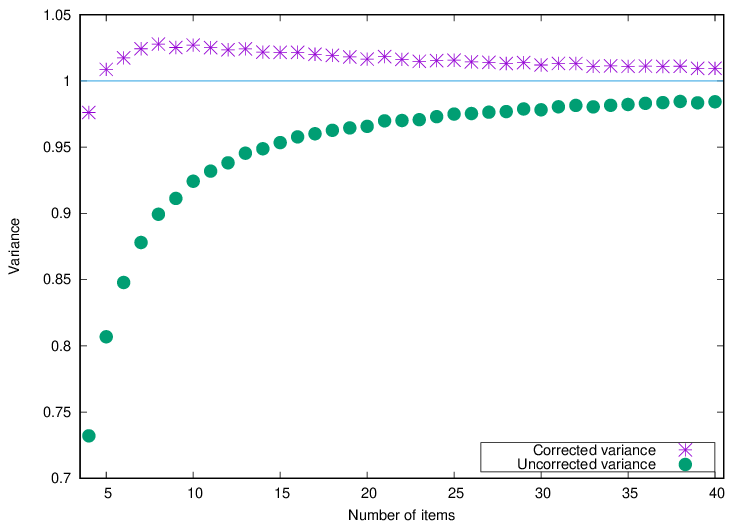}}
\caption{\label{figd}Estimated variance of loadings and corrected variance by number of items when true variance is unity.}
\end{figure}

The estimator for $\hat{\tau}^2$ is biased downwards, and $\hat{\tau}^2$ 
is zero for a two-item scale. The bias depends on the distribution of 
the loadings. For an exponential distribution, the correction $\hat{\tau^2}=k\{(\bar{c^2})^{1/2}-\bar{c}\}/(k-2)$
works very well. However, from examining loadings, it seems that their 
distribution is usually roughly normal. It seems from simulations that 
the smaller correction $\hat{\tau}^2=k\{(\bar{c^2})^{1/2}-\bar{c}\}/(k-1)$ 
works best Figure \ref{figd} shows the variance and corrected variance 
average from 100000 simulations from normally-distributed loadings, where 
the true variance is unity.

The confidence interval on $\alpha_m$ could be estimated by resampling 
$x_{ij}$ from ${\bf V}$ assuming multivariate normality. This can be done 
using a Cholesky decomposition of ${\bf V}$. The resampled ${\bf V}$ matrices 
can be used to derive resampled reliabilities. This 
bootstrap method can yield a good confidence interval for $R$.

\section{Inference for the reliability}
This is not often discussed, as factor-analysis software is used. We 
show how the common-factor and 2-factor models can be fitted using more 
general software, in a way that is computationally quicker.
More importantly, it allows computation of the confidence interval for reliability, and
facilitates calculation of it when some items are missing.

Inference proceeds by maximising the log-likelihood for the $2k$ parameters
$\sigma^2_1\cdots \sigma^2_k$ and $\lambda_1\cdots\lambda_k$. A reparameterization
is used here so that the reliability itself becomes one of the parameters.
A function minimiser that computes standard errors will then deliver 
the reliability and its standard error. Another consequence is that the 
reliability parameter can be fixed at different values to enable the 95\% 
confidence interval on $R$ to be found from the profile likelihood. Fortunately,
the log-likelihood can be written down simply, facilitating this programme.
\subsection{The likelihood function}
The likelihood of obtaining the observed results for the $i$th individual
is  under the model
\[\mathcal{L}_i=\frac{\exp\{-{\bf x}_i^T{\bf V}^{-1}{\bf x}_i/2\}}{(2\pi)^{k/2}|{\bf V}|^{1/2}},\]
where ${\bf x}_i$ denotes the vector $(x_{i1} \ldots x_{ik})$ \etc and
${\bf V}$ is the covariance matrix. The full likelihood is $\mathcal{L}=\prod_{i=1}^n
\mathcal{L}_i$.

Fortunately, the inverse matrix ${\bf V}^{-1}$ and the determinant $|{\bf
V}|$ can be evaluated as functions of the $\sigma_j^2$ and $\lambda_j$.
The matrix ${\bf V}={\bf D}+ {\boldsymbol \lambda}{\boldsymbol \lambda}^T$,
where ${\bf D}$ is the diagonal matrix with elements $D_{jj}=\sigma^2_j$.

From the Sherman-Morrison formula (see \eg Press {\em et al} 2007) we
have that
\[{\bf V}^{-1}={\bf D}^{-1}-{\boldsymbol \lambda}{\boldsymbol \lambda}^T /(1+{\boldsymbol \lambda}^T){\bf D}^{-1}{\boldsymbol \lambda}.\]
The determinant
\[|{\bf V}|=1+{\boldsymbol \lambda}^T{\bf D}^{-1}{\boldsymbol \lambda}=1+\sum_{j=1}^k \lambda_j^2/\sigma_j^2.\]

This yields the log-likelihood function
\[\ell=(-1/2)\sum_{i=1}^n\sum_{j=1}^k x_{ij}^2u_j +(1/2) T/(1+\sum_{j=1}^k \lambda_j^2u_j)+(n/2)\sum_{j=1}^k \ln (u_j)-(n/2)\ln(1+\sum_{j=1}^k \lambda_j^2u_j)-(nk/2)\ln(2\pi),\]
where  $u_j=1/\sigma_j^2$, and $T=\sum_{i=1}^n (\sum_{j=1}^k x_{ij}\lambda_j u_j)^2$.

The log-likelihood can also be derived from the likelihood of the observed
data conditional on the true (latent) score $\theta$. One then integrates
over the standard normal distribution of $\theta$. The latent variable 
approach is given in Bartholemew {\em et al} (2011).

It can be seen that Mcdonald's omega can cope with missing items. One
computes the relevant ${\bf V}_i$ for the $i$th individual, omitting 
the missing items. Cronbach's alpha can also cope with missing items. 
None of this requires sophisticated techniques such as multiple imputation.
\subsection{Reparameterization}
We want one of the $2k$ parameters to be the reliability $R$. This can
be achieved in several ways (such as using hyperspherical coordinates)
but the following method is the simplest. We take parameters 1 to $k$
to be the $u_j$, then $k=1$ parameters $z_1  \cdots z_{k-1}$ and finally
$R$. The conversion to the $\lambda_j$ is as follows: with $S=\sum_{j=1}^k
\sigma_j^2$,
\[\lambda_j=\frac{\sqrt{S}(R/(1-R))z_j}{\sum_{j=1}^{k-1}z_j+1},\]
for $j$ from 1 to $k-1$, and finally
\[\lambda_k=\frac{\sqrt{S}(R/(1-R))}{\sum_{j=1}^{k-1}z_j+1}.\]
This transformation ensures the correct formula for $R$. We take the 
$z_j > 0$ and $R > 0$ which ensures $\lambda_j > 0$.

For starting values, we take the $\sigma_j^2$ to be the sample variances,
${\bf z}={\bf 1}$ and $R=0.5$.
\subsection{Goodness of fit: Comparison with saturated model}
In the saturated model, the covariance matrix is estimated as the sample
covariance. The log-likelihood becomes
\[\ell_s=-nk/2-(n/2)\ln(|\hat{\bf V}|)-(nk/2)\ln(2\pi),\]
where $|\hat{\bf V}|$ is evaluated by doing a Cholesky decomposition 
and taking the squared product of the diagonal elements of ${\bf L}$, 
where ${\bf V}={\bf L}{\bf L}^T$, where ${\bf L}$ is a lower triangular
matrix. This enables a goodness of fit test for the model when $k > 3$,
where the chi-squared $X^2=2(\ell_s-\ell)$ has $k(k-3)/2$ degrees of 
freedom. Note that the saturated model does not yield an estimate of reliability.

This model gave really bad fits for all but one of the datasets examined,
and $|{\bf V}|$ is very sensitive to errors. It was thought better to 
fit a 2-factor model and derive a goodness of fit statistic from that. 
This model is described in appendix A.
\
\section{Data fitting}
\subsection{Simulation}
Table \ref{tabc} is based on 10000 simulations of the common factor model 
from different reliabilities, number of cases and number of items. The 
bias is always very small, much smaller than the statistical standard 
error of the estimated reliability.

It is usually negative (the estimate is too low) but sometimes becomes 
positive at low reliability. The standard error increases with lower reliability.
The conclusion is that bias of the maximum likelihood estimator is not 
a problem.

The table also shows Cronbach's alpha, and the modified version given 
earlier. It can be seen that the modified version gives reliabilities 
close to the correct value, with standard error comparable to the maximum-likelihood 
estimate of McDonald's omega.
\begin{table}[h]
\begin{tabular}{|c|c|c|c|c|c|c|}  \hline
No. cases $n$&no. items $k$&$R$&bias&s.e.&$\alpha$ & $\alpha_m$ \\ \hline
100&4&0.9&-0.0018&0.0229&0.784 (0.091)&0.886 (0.042)\\ \hline
100&4&0.7&0.0021&0.0594&0.605 (0.090)&0.692 (0.059)\\ \hline
100&4&0.4& 0.0448&0.0873&0.337 (0.115)&0.420 (0.088)\\ \hline
100&15&0.9&-0.008&0.0211&0.846 (0.030)&0.893 (0.019) \\ \hline
100&15&0.7& 0.004&0.0417&0.653 (0.052)&0.699 (0.044)\\ \hline
100&15&0.4&0.0282&0.0719&0.363 (0.094)&0.426 (0.080)\\ \hline
1000&10&0.9&-0.0002&0.0125&0.828 (0.048)&0.887 (0.024)\\ \hline
1000&10&0.7&0.0004&0.0139&0.643 (0.0346)&0.691 (0.023) \\ \hline
1000&10&0.4&0.0031&0.0261&0.367 (0.0345)&0.398 (0.029)\\ \hline
\end{tabular}
\caption{\label{tabc}Simulation results: true reliability, bias of estimates, standard error of reliability, Cronbach's alpha and $\alpha_m$.}
\end{table}
\subsection{Fitting real datasets}
To illustrate the methods of computation proposed, some datasets were
analysed, described in table \ref{taba}. Some of them were not used for
their intended purpose, and explanatory variables were discarded.

Results of the model fitting are given in table \ref{tabb}. Several findings
emerge. Cronbach's alpha and the modified version often give a very similar 
result to McDonald's omega. The reliability $R_i$ indeed gives higher 
values than $R$, and weighted estimates of the score should be used where 
possible.

The goodness of fit test shows that when there are more than very few
variables, the 1-factor model cannot capture the detail of the covariance
matrix, and at least 2 factors are needed. The model fits badly, and 
with a lot of cases, this becomes very evident. The reliability calculated 
under the 2-factor model is sometimes lower and sometimes higher than 
for the 1-factor model. Under the 2-factor model there is an additional 
term in the denominator of the reliability, but the variances $\sigma_j^2$ 
will be smaller.

\begin{figure}
\centering
\makebox{\includegraphics{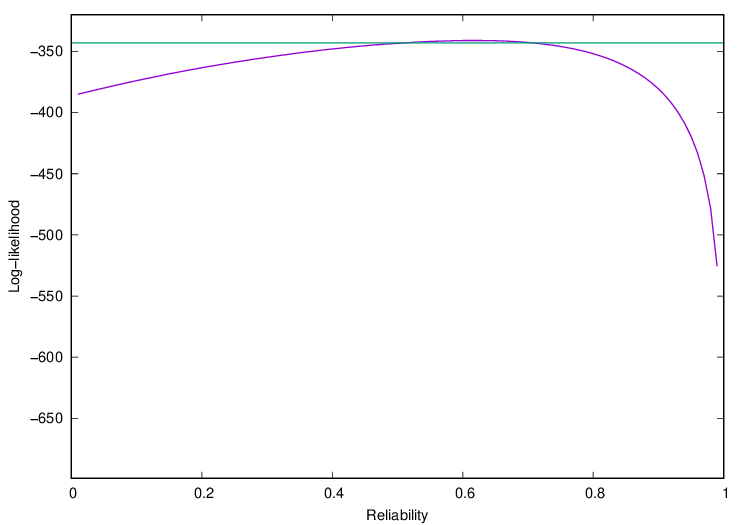}}
\caption{\label{figc}Profile log-likelihood by reliability for the happiness dataset. The limits of the 95\% confidence interval
are where the horizontal line 1.92 below the maximum log-likelihood cuts the log-likelihood curve.}
\end{figure}

Figure \ref{figc} shows the profile likelihood for the happiness dataset
as a function of reliability, and the 95\% confidence interval. It can
be seen that this interval is larger in the direction of lower reliability.
This means that uncertainty about the exact value of the reliability 
would reduce the value of the test further.
\begin{table}[h]
\begin{tabular}{|l|l|c|c|l|}  \hline
Dataset&Description&no. cases&no. vars&source \\ \hline
Happy&Happiness survey&143&6&Kaggle \\ \hline
Airline&Satisfaction survey&129880&14&Kaggle \\ \hline
Tech&Enthusiasm for new technology&107&29&Kaggle \\ \hline
Stress&Stress among students survey&843&23&Kaggle \\ \hline
Customer&Customer satisfaction at Kashmir cafe&10616&4&Kaggle \\ \hline
Website&Satisfaction with an e-commerce website&73&26&Kaggle \\ \hline
Nerdy&Nerdy personality attributes scale&25459&26&Psychometrics project \\ \hline
\end{tabular}
\caption{\label{taba}Brief description of the datasets used in the analyses.The last one is from the Open source psychometrics project}
\end{table}
\begin{table}[h]
\begin{tabular}{|l|c|c|c|c|c|c|c|c|}  \hline
Dataset&Alpha&$\alpha_m$&Omega (95\% CI.)& $R_i$&$R_2$&CV & $X^2$[df]&p-value \\ \hline
Happy &0.606&0.626&0.618 (0.511,0.705)&0.690&0.609 & 0.590 & 15.6 [6] & 0.0514 \\ \hline
Airline&0.772&0.787&0.725 (0.722,0.727)&0.892&0.628&0.796&4176354 [14]&0 \\ \hline
Tech&0.857&0.871&0.803 (0.736,0.855)&0.929&0.803&0.597&368.1 [29]& 0 \\ \hline
Stress&0.697&0.711&0.693 (0.662,0.723)&0.711&0.705&0.358 &582.5 [23]& 0 \\ \hline
Customer& 0.633&0.733&0.736 (0.727,0.744)&0.843&0.731 & 0.881&4.4 [4] & 0.35 \\ \hline
Website&0.959&0.961&0.960 (0.945,0.972)&0.969&0.870&0.292&183.5 [26] & 0 \\ \hline
Nerdy&0.850&0.856&0.856 (0.845,0.851)&0.868&0.783&0.266&28276 [26] &0\\ \hline
\end{tabular}
\caption{\label{tabb}Cronbach's alpha, modified Cronbach's alpha, McDonald's omega with 95\% confidence interval, $R_i$ computed weighting scores inversely by variance,
$R_2$ from the 2-factor model, coefficient of variation of the $\lambda_j$ 
and goodness of fit chisquared and p-value.}
\end{table}
\section{Conclusions}
The cost functions derived here quantify the consequences of low reliability
for several decision-relevant metrics. These results indicate that, for 
classification purposes, reliabilities above 0.7 are often necessary.

Efficient likelihood-based procedures for estimating reliability under 
common-factor and two-factor models were also presented, together with 
a profile-likelihood confidence interval that performs well in practice.
An advantage of this approach is that missing values are easily accommodated.
When item weights can be estimated, forming a weighted score reduces 
the variance of the latent-trait estimator and therefore increases reliability.

A modified Cronbach's alpha was also proposed. It agrees closely with 
McDonald's omega when the common-factor model is correct and has a similar 
standard error, yet it requires no model fitting. This makes it particularly 
useful in small samples where omega is difficult to compute reliably.

Further work could examine these cost functions in more depth. Item Response 
Theory (IRT) is another methodology that is attracting more interest (\eg 
Tutz, 2025). This produces a reliability that is a function of the true 
score $\theta$. Extending the cost-based perspective to that setting would 
also be worthwhile.

\section*{Appendix A: the 2-factor model}
The model is $x_{ij}=\sigma_j\epsilon_j+\lambda_j\theta+\eta_j \phi$,
where $\theta$ and $\phi$ are the two normally-distributed latent variables.
The reliability is then
\[R_2=\frac{(\sum_{j=1}^k \lambda_j)^2}{\sum_{j=1}^k \sigma_i^2+(\sum_{j=1}^k \lambda_j)^2+(\sum_{j=1}^k \eta_j)^2}.\]
In general, one can rely on matrix algebra or integration to derive the 
likelihood function. Here we use integration. The contribution to the 
likelihood from the $i$th individual is
\[{\mathcal L}_i=\frac{1}{(2\pi)^{k/2+1}}\int_{-\infty}^\infty\int_{-\infty}^\infty \exp[-(1/2)\{\frac{\sum_{j=1}^k (x_{ij}-\lambda_j\theta)^2}{\sigma_j^2}+\theta^2+\phi^2\}]\dee \theta\dee \phi.\]
The double integral can be evaluated by completing the squares in the 
exponent for $\theta$ and $\phi$, and using the properties of the multivariate 
normal distribution. After some algebra,
\begin{eqnarray*}\ell&=&-nk\ln(2\pi)/2-n/2\sum_{j=1}^k \ln \sigma_j^2 -(n/2)\ln(AB-E^2)-\sum_{i=1}^n\sum_{j=1}^k x_{ij}^2/\sigma_j^2-(n/2)\ln(AB-E^2)+\\
&&(1/2)\sum_{i=1}^n\frac{AX_i^2+BY_i^2+2EX_iY_i}{(AB-E^2)^2},\end{eqnarray*}
where
\[A=\sum_{j=1}^k \lambda_j^2/\sigma_j^2+1,\]
\[B=\sum_{j=1}^k \eta_j^2/\sigma_j^2+1,\]
\[C_i=\sum_{j=1}^k x_{ij}\lambda_j/\sigma_j^2,\]
\[D_i=\sum_{j=1}^k x_{ij}\eta_j/\sigma_j^2,\]
\[E=\sum_{j=1}^k \lambda_i\eta_i/\sigma_i^2,\]
\[X_i=ED_i-BC_i,\]
\[Y_i=EC_i-AD_i.\]
\end{document}